\documentclass[12pt]{article}
\usepackage[T1]{fontenc}
\usepackage{amsfonts}
\usepackage{latexsym}
\usepackage{dsfont}
\usepackage{amsmath}
\usepackage{amssymb}
\usepackage{amssymb}
\usepackage{amsthm}
\usepackage{mathrsfs}
\usepackage[nosort]{cite}
\usepackage{setspace}
\usepackage{color}
\usepackage{graphicx}
\usepackage{accents}
\usepackage[font=footnotesize,labelfont=bf]{caption}
\usepackage[utf8]{inputenc}
\usepackage{geometry}
\usepackage[toc,page]{appendix}

\usepackage{tikz}
\usetikzlibrary{arrows.meta,backgrounds,calc,fit,positioning,scopes,shadows}

\makeatletter
\def\tikzsavelastnodename#1{\let#1=\tikz@last@fig@name}
\makeatother

\usepackage[hidelinks]{hyperref}
\hypersetup{
colorlinks=false,
citecolor= blue,
linkcolor= blue,
urlcolor= blue,
breaklinks=true
}
\usepackage{cleveref}
\crefformat{section}{\S#2#1#3} % see manual of cleveref, section 8.2.1
\crefformat{subsection}{\S#2#1#3}
\crefformat{subsubsection}{\S#2#1#3}

\def\appendix#1{\addtocounter{section}{1}\setcounter{equation}{0}
\renewcommand{\thesection}{\Alph{section}}
\section*{Appendix \thesection\protect\indent \parbox[t]{11.15cm}{#1}}
\addcontentsline{toc}{section}{Appendix \thesection\ \ \ #1}}

\numberwithin{equation}{section}

\makeatletter
 \let\old@startsection=\@startsection
 \let\oldl@section=\l@section
 \renewcommand{\@startsection}[6]{\old@startsection{#1}{#2}{#3}{#4}{#5}{#6\mathversion{bold}}}
 \renewcommand{\l@section}[2]{\oldl@section{\mathversion{bold}#1}{#2}}
\makeatother

\def\dd{\text{d}}

\begin{document}

%%%%%%%%%%%%% TITLE %%%%%%%%%%%%%%%%%%%%%%%%%

\begin{titlepage}
\vspace*{-1.0cm}

\begin{center}
%\today

\hfill {\footnotesize ZMP-HH/24-29}%\\
%\hspace{13.21cm}TCDMATH-XX}

\vspace{2.0cm}

{\LARGE  {\fontfamily{lmodern}\selectfont \bf Revisiting the Symmetries of  \\
\vspace{2mm}
Galilean Electrodynamics}} \\[.2cm]
\normalsize

\vskip 1.5cm
\textsc{Andrea Fontanella$^{\mathfrak{Re}}$ \footnotesize and \normalsize Juan Miguel Nieto Garc\'ia$^{\mathfrak{Im}}$}\\
\vskip 1.2cm

\begin{small}
{}$^{\mathfrak{Re}}$ \textit{School of Mathematics $\&$ Hamilton Mathematics Institute, \\
Trinity College Dublin, Ireland}\\
\vspace{1mm}
\href{mailto:andrea.fontanella[at]tcd.ie}{\texttt{andrea.fontanella[at]tcd.ie}}

\vspace{5mm}
{}$^{\mathfrak{Im}}$ \textit{II. Institut für Theoretische Physik, Universität Hamburg,\\
Luruper Chaussee 149, 22761 Hamburg, Germany} \\
\vspace{1mm}
\href{mailto:juan.miguel.nieto.garcia[at]desy.de}{\texttt{juan.miguel.nieto.garcia[at]desy.de}}

\end{small}

\end{center}

\vskip 1 cm
\begin{abstract}
\vskip1cm

\noindent We determine the symmetries of four different theories: I) Galilean Electrodynamics (GED), II) GED coupled to 5 free static scalar fields, III) Galilean Yang-Mills (GYM), and IV) GYM coupled to 5 interacting scalar fields. We correct some old results in the literature, by finding that the symmetries of GED in a spacetime of generic dimension $d+1$ is always infinite dimensional, and in $3+1$ they correspond to the conformal Milne algebra extended by a spatial dilatation generator, which we call $D_x$. Finally, we comment on how these results fit into the framework of the non-relativistic AdS$_5$/CFT$_4$ correspondence.

\end{abstract}

\end{titlepage}

\tableofcontents
\vspace{5mm}
\hrule

%%%%%%%%%%%%%%%%%%% BODY %%%%%%%%%%%%%%%%%%

\setcounter{section}{0}
\setcounter{footnote}{0}

\section*{Introduction}

Galilean Electrodynamics (GED) is a very important and simple theory, so that it may be regarded as the ``hydrogen atom'' of non-relativistic field theories. This theory was first proposed in \cite{Santos:2004pq} as a Lagrangian formulation for the electric and magnetic non-relativistic limits of the Maxwell's equations proposed by Le Bellac and L\'evy-Leblond in \cite{LeBellac}, as Galilean invariant equations of electromagnetisms. 
Later, GED was also derived as a non-relativistic limit of the Maxwell theory with an additional scalar field \cite{Bergshoeff:2015sic}. The non-abelian generalisation, called Galilean Yang-Mills (GYM), was proposed in \cite{Bagchi:2015qcw}. 
Both GED and GYM theories, opportunely supplemented by five additional scalar fields, also appear when considering the gauge theory dual of the GGK non-relativistic string theory with target space the String Newton-Cartan version of AdS$_5\times$S$^5$ \cite{Gomis:2005pg, Fontanella:2024kyl, Fontanella:2024rvn}. The study of the non-relativistic AdS/CFT correspondence is currently a growing field, see \cite{Lambert:2024uue, Lambert:2024yjk, Lambert:2024ncn, Blair:2024aqz} for other recent work on the subject.  

Regarding the identification of the symmetries of GED, there has been some confusion in the past, which we wish to clarify in this article. 
Symmetries of the GED theory have been analysed in \cite{Festuccia:2016caf}, and they have been distinguished between symmetries of the equations of motion (named ``on-shell'') and symmetries of the action (named ``off-shell''). The on-shell symmetries were found to be infinitely many, with the peculiarity that in 3+1 dimensions they contain a finite truncation of the Galilean Conformal Algebra (GCA). On the other hand, the off-shell symmetries were found to be finite, and with the property of containing only a Lifshitz scale invariance instead of the finite truncation of the GCA. However, the reason for finding only a finite amount of off-shell symmetries can be traced back to the fact that in \cite{Festuccia:2016caf} the Lagrangian was demanded to transform under the symmetries as a total derivative of a specific type, namely $\delta\mathcal{L}=\partial_\mu\left(\xi^\mu\mathcal{L}\right)$, instead of transforming as $\delta \mathcal{L} = \partial_{\mu} F^{\mu}$, where $F^{\mu}$ is a generic function of fields, coordinates and parameters, and $\xi^\mu$ is a vector field containing the transformation parameters. As it is standard practice in QFT, any boundary term generated by the symmetry can be eliminated by imposing suitable boundary conditions, e.g. the fall-off condition on the fields at infinite spatial distance.

In the more recent paper \cite{Bagchi:2022twx}, the symmetries of the GED and GYM theories were considered again. In 2+1 dimensions, they found that they are infinitely many, and we do match their result. However, in 3+1 dimensions, i.e. the case relevant for holography, they obtain that the GED action admits the full GCA symmetry. As we shall comment in section \ref{sec:GED_symmetries}, we do not reproduce this result.\footnote{We believe in \cite{Bagchi:2022twx} there was a misuse of the name ``Galilean Electrodynamics'' for the case in 3+1 dimensions. In particular, we believe they actually refer to a non-relativistic limit of the Maxwell theory with no extra scalar field, where the symmetries were found to be the full GCA in the older paper \cite{Bagchi:2014ysa}.} Instead, we find that the symmetries of the GED theory in 3+1 dimensions are still infinitely many, but they correspond to a truncation of the GCA augmented by an additional generator.  

The plan of this article is the following. In section \ref{sec:GED_symmetries} we systematically revisit the symmetries of the GED theory in any spacetime dimension. We impose that the Lagrangian transforms as a total derivative of a generic function and, in this way, we find that the symmetries are infinitely many. For a case of particular interest, namely $d=3$, we show that they correspond to the conformal Milne algebra extended by a spatial dilatation generator that we call $D_x$. In section \ref{ansatz_GED+scalars}, we determine the symmetries of the GED action with 5 free static scalars that appear from taking the non-relativistic limit of the DBI action studied in \cite{Fontanella:2024kyl, Fontanella:2024rvn}. In section \ref{sec:GYM_symmetries} we revisit the symmetries of GYM, and we find that they are given by the conformal Milne algebra, confirming the result of \cite{Bagchi:2022twx}. In section \ref{sec:GYM_5scalars_symmetries} we determine the symmetries of the GYM theory coupled to 5 interacting adjoint scalar fields which appears from the non-relativistic limit of the DBI action, as found in \cite{Fontanella:2024kyl, Fontanella:2024rvn}. In section \ref{sec:application_NR_holography} we will compare the symmetries obtained in this work with the Killing vector symmetries of the GGK theory found in \cite{Fontanella:2024kyl, Fontanella:2024rvn}. This leads us to identify GYM with 5 scalars as the most promising candidate for its dual gauge theory.

\section{Symmetries of the GED action}
\label{sec:GED_symmetries}
We shall use the same notation of \cite{Festuccia:2016caf}, where the GED action in flat $d+1$ dimensions is 
\begin{equation}\label{GED}
S_{\text{GED}}=\int \dd^{d+1}x\left[-\frac{1}{4}f^{ij}f_{ij}- E^i\partial_i\varphi+\frac{1}{2}\left(\partial_t\varphi\right)^2\right]\,.
\end{equation}
where $f_{ij}$ and $E_i$ are defined as
\begin{equation}
f_{ij}\equiv \partial_i a_j-\partial_j a_i \, , \qquad
E_i\equiv -\partial_i\tilde\varphi-\partial_ta_i \, .
\end{equation}
The spacetime coordinates are indicated by $(t, x^i)$,  $a_i$ is a vector potential with only spatial components, $\varphi$ and $\tilde \varphi$ are scalar fields. All fields have a full dependence on the spacetime coordinates. 

We want to determine the symmetries of the GED action, which are transformations of the spacetime coordinates and of the fields. We denote the transformations of the spacetime coordinates by
\begin{eqnarray}
    \delta t = \xi^t \, , \qquad
    \delta x^i = \xi^i \, .
\end{eqnarray}
and we allow the fields to transform with the most generic linear ansatz
\begin{subequations}\label{ansatz}
\begin{align}
\delta\varphi & =  \xi^t\partial_t\varphi+\xi^k\partial_k\varphi + \alpha_1 \varphi + \alpha_2 \tilde{\varphi} + \beta^k a_k \,,\\
\delta\tilde\varphi & =  \xi^t\partial_t\tilde\varphi+\xi^k\partial_k\tilde\varphi + \alpha_3 \varphi + \alpha_4 \tilde{\varphi} + \gamma^k a_k \,,\\
\delta a_i & = \xi^t\partial_t a_i+\xi^k\partial_k a_i + \zeta_i \varphi + \varepsilon_i \tilde{\varphi} + \pi_i{}^k a_k \,,
 \end{align}
\end{subequations}
where $\xi^t, \xi^i, \alpha_i, \beta_i, \gamma_i, \zeta_i, \varepsilon_i, \pi_{ij}$ are all functions of the spacetime coordinates. We demand that the GED Lagrangian transforms under (\ref{ansatz}) as a total derivative, i.e. $\delta \mathcal{L} = \partial_{\mu} F^{\mu}$, where $F^{\mu}$ is a generic function of fields, coordinates and parameters. This fixes  the vector field components $\xi^t$ and $\xi^i$ to
\begin{subequations}\label{Xi_vector}
\begin{align}
\xi^t &= - \frac{d-4}{2} \beta t^2 + \gamma t + \zeta \, , \\
\xi^i &= b_i(t) + \alpha x^i + \beta t x^i +  \lambda^i{}_k x^k \, ,   
 \end{align}
\end{subequations}
where $\alpha, \beta, \gamma, \zeta, \lambda_{ij}$ are constant parameters, with $\lambda_{ij} = - \lambda_{ji}$, and $b_i(t)$ are generic functions of $t$. The functions $\alpha_i, \beta_i, \gamma_i, \zeta_i, \varepsilon_i, \pi_{ij}$ are fixed in terms of derivatives of $\xi^t$ and $\xi^i$, which gives us
\begin{subequations}\label{func_fix}
\begin{align}
\alpha_1 & = \frac{d}{2}\alpha + (d-2) \beta t - \frac{\gamma}{2} \, , &
\alpha_4 &= \frac{d-4}{2}\alpha - (d-4) \beta t + \frac{3}{2}\gamma   \, ,  \\
\gamma_i &= \zeta_i = - \beta x_i - \partial_t b_i \, , & 
\pi_{ij}& = \partial_j \xi_i +  \left(\frac{1}{2}\gamma + \frac{d-4}{2} \alpha \right) \delta_{ij} \, , \\
\alpha_2 &= \alpha_3 = \beta_i = \varepsilon_i = 0 \, . 
 \end{align}
\end{subequations}
The symmetries of the GED action generate an infinite dimensional algebra, spanned by the generators associated with the transformation parameters $\alpha, \beta, \gamma, \zeta, \lambda_{ij}, b_i (t)$. They are:\footnote{Our $D_t$ and $D_x$ correspond to $D_1$ and $D_2$ of \cite{Festuccia:2016caf} respectively.}
\begin{subequations}\label{GED_generators}
\begin{align}
\alpha&:  &D_x=x^i\partial_i\, , \\
\beta&: &K=-\frac{1}{2}(d-4)t^2\partial_t+tx^i\partial_i\, ,\\
\gamma&:  &D_t=t\partial_t\, ,\\
\zeta&: &H=\partial_t\, ,\\
\lambda_{ij}&: &J_{ij}=x^i\partial_j-x^j\partial_i\, ,\\
b_i (t)&: &M^{(n)}_i =t^{n+1}\partial_i\, . 
 \end{align}
\end{subequations}
The non-vanishing commutation relations are:
\begin{subequations}\label{GED_symm_algebra}
\begin{align}
[D_t\,,H]&=-H \, , & 
[K\,,M^{(n)}_i] &=-\frac{1}{2} \left[d-2+(d-4)n\right]M^{(n+1)}_i \, , \\
[D_t\,,M^{(n)}_i]&=(n+1)M^{(n)}_i \, , & [D_x\, ,  M^{(n)}_i] &= - M^{(n)}_i  \, ,   \\
[D_t\,,K]&=K \, ,  & \\
[H\,,M^{(n)}_i] &=(n+1)M^{(n-1)}_i  \, , & [K\,,H]&=(d-4)D_t-D_x \, ,   \\
[ J_{ij}\,, J_{k\ell} ] &= 2 \delta_{k[j} J_{i]\ell} - 2 \delta_{\ell[j} J_{i]k} \, ,  & [J_{ij}\,,M^{(n)}_k]&=\delta_{jk} M^{(n)}_i - \delta_{ik} M^{(n)}_j \,  . 
 \end{align}
\end{subequations}
where we use the convention $2 A_{[i} B_{j]} \equiv A_i B_j - A_j B_i$. Notice that, for $d\neq 4$, we can define $D\equiv D_t - \frac{1}{d-4} D_x$, which behaves like a usual dilatation operator. 

The $d=3$ case is special, because this is the only dimension where the generator $K$ gains meaning of special conformal transformation and $D$ gains meaning of an actual dilatation transformation.
It is easy to check that in $d=3$ the algebra (\ref{GED_symm_algebra}), without the generator $D_x$, is a truncation of GCA. The GCA, first introduced in \cite{Bagchi:2009my}, is spanned by 
\begin{eqnarray}\label{GCA_generators}
    L^{(n)}=t^{n+1}\partial_{t} +(n+1)t^{n}x^{i}\partial_{i} \, , \qquad\quad
	M_{i}^{(n)} =t^{n+1}\partial_{i} \, ,  \qquad
    J_{ij}=x^i\partial_j-x^j\partial_i\, ,
\end{eqnarray}
where $L^{(n)}$ corresponds to $H, D, K$ for $n=-1, 0, 1$ respectively. It is important to note that the full tower of generators $L^{(n)}$ is not a symmetry of the GED action in $d=3$. By requiring that the whole tower of generators $L^{(n)}$ transforms the GED Lagrangian as a total derivative of a generic function, one gets the condition that $n$ must be equal to either $-1,0,1$. Therefore, only the finite truncation of $L^{(n)}$ that corresponds to $H, D$ and $K$ is a symmetry of the GED theory in $d=3$. The algebra generated by $\{ H, J_{ij}, M_{i}^{(n)} \}$ is also called the Milne algebra \cite{Duval:1993pe}.\footnote{We thank A. Bagchi for pointing this out.} Therefore, the symmetries of the GED action in $d=3$ are given by a \emph{Conformal Milne Algebra}, augmented by the spatial dilatation generator $D_x$.     

We conclude this section with a comment on the parameter $\alpha_3$ entering the field transformation (\ref{ansatz}) that was already noted in \cite{Festuccia:2016caf}. This parameter needs to be zero in order to leave the GED Lagrangian invariant up to a total derivative. However, if we set $\alpha_3 = \alpha_3(t)$, i.e. a generic function of $t$, this would not be a symmetry of the action, but it would leave the equations of motion invariant. More precisely, this transformation generates terms in the action which are proportional to the equations of motion, and therefore it is a symmetry only on-shell.

\section{Symmetries of the GED action with 5 free static scalars}
\label{sec:GED_5scalars__symmetries}

One of the non-relativistic limits of the $\mathcal{N}=4$ Yang-Mills action proposed in \cite{Fontanella:2024kyl, Fontanella:2024rvn} is ($N^2$ copies of) GED theory with five extra free static scalar fields. In this section, we aim to identify the symmetries of this theory. 

The GED with scalars action in $d=3$ is the following
\begin{eqnarray}
    S = S_{\text{GED}} + S_{\text{scalars}} = \int \dd t \dd^{3}x \left[-\frac{1}{4}f^{ij}f_{ij}- E^i\partial_i\varphi+\frac{1}{2}\left(\partial_t\varphi\right)^2 - \frac{1}{2} (\partial_i S^I )^2 \right]\,.
\end{eqnarray}
where $S^I$, with $I=1,..., 5$, are the five scalars.
We can now ask whether the scalars $S^I$ break any of the GED symmetries. To answer this question, we first write down the most general linear ansatz
\begin{subequations}\label{ansatz_GED+scalars}
\begin{align}
\label{dphi_GED+scalars}
\delta\varphi & =  \xi^t\partial_t\varphi+\xi^k\partial_k\varphi + \alpha_1 \varphi + \alpha_2 \tilde{\varphi} + \beta^k a_k + \tau_I S^I \,,\\
\delta\tilde\varphi & =  \xi^t\partial_t\tilde\varphi+\xi^k\partial_k\tilde\varphi + \alpha_3 \varphi + \alpha_4 \tilde{\varphi} + \gamma^k a_k + \sigma_I S^I\,,\\
\delta a_i & = \xi^t\partial_t a_i+\xi^k\partial_k a_i + \zeta_i \varphi + \varepsilon_i \tilde{\varphi} + \pi_i{}^k a_k + \chi_{iI} S^I\,, \\
\delta S^I & = \xi^t\partial_t S^I+\xi^k\partial_k S^I + \mu^I \varphi + \nu^I \tilde{\varphi} + \rho^{I k} a_k + \omega^I\null_{J} S^J \, , 
 \end{align}
\end{subequations}
where $\xi^t, \xi^i, \alpha_i, \beta_i, \gamma_i, \zeta_i, \varepsilon_i, \pi_{ij}$ are fixed accordingly to equations (\ref{Xi_vector}) and (\ref{func_fix}), namely we demand that the GED symmetries are preserved. To fix the remaining functions $\mu^I, \nu^I,\rho^{I i},\omega^I\null_{J}, \tau_I, \sigma_I, \chi_{iI}$, we demand that under (\ref{ansatz_GED+scalars}) the Lagrangian transforms as a total derivative of a generic function of fields, coordinates and parameters. This imposes the following restrictions,\footnote{The indices $I, J,\dots$ are raised and lowered with the $\delta_{IJ}$.} 
\begin{subequations}\label{fix_func_2}
\begin{align}
\mu_I &= - \sigma_I = f_I(t) \, , \qquad
\omega_{IJ} = \tilde{\omega}_{IJ} (t) +  \frac{1}{2} \left(\alpha + \gamma + 2 \beta t  \right) \delta_{IJ} \, , \\
\nu_I &= \tau_I = \chi_{iI} = \rho^{I i} = 0 \, ,
 \end{align}
\end{subequations}
where $\tilde{\omega}_{IJ}(t)$ are generic functions of time and antisymmetric, $\tilde{\omega}_{IJ} (t) = - \tilde{\omega}_{JI} (t) $, and $f_I(t)$ are generic functions of $t$. 
From this, we learn that the five additional scalars preserve the GED symmetries. The term in the variations associated with the diagonal part of $\omega_{IJ}$ is an internal transformation of the scalar fields $S^I$, which is fixed to precisely compensate extra terms generated by the spacetime symmetries of the GED action coming from the parameters $\alpha, \beta$ and $\gamma$.  
The terms associated with $\mu_I$ and the off-diagonal part of $\omega_{IJ}$ are instead purely internal symmetries. They generate an infinite dimensional algebra, spanned by the following generators,
\begin{subequations}\label{internal_generators}
\begin{align}
f_I(t)&:  &J^{(n)}_{0I} = t^n \varphi \frac{\partial}{\partial S^I} - t^n S^I \frac{\partial}{\partial S^0} \, , \\
\tilde{\omega}_{IJ}(t)&: &J_{IJ}^{(n)} = t^n S^I \frac{\partial}{\partial S^J} - t^n S^J \frac{\partial}{\partial S^I} \, . 
 \end{align}
\end{subequations}
where we denoted $S^0 \equiv \tilde{\varphi}$, which has meaning of auxiliary direction. The non-vanishing commutation relations are
\begin{align} \label{comm_rotations_GED}
\null [ J_{IJ}^{(n)}, J_{KL}^{(m)} ] &= 2 \delta_{K[J} J_{I]L}^{(n+m)} - 2 \delta_{L[J} J_{I]K}^{(n+m)} \, , & 
\null [J_{IJ}^{(n)}, J_{0K}^{(m)}] &= \delta_{JK} J_{0I}^{(n+m)} - \delta_{IK} J_{0J}^{(n+m)} \, ,
 \end{align}
 together with
\begin{subequations}\label{comm_HDK_rotations_GED}
\begin{align}
\null [H, J_{0I}^{(n)}] &= n J_{0I}^{(n-1)} \, , &\qquad 
\null [D, J_{0I}^{(n)}] &=  n J_{0I}^{(n)} \, , &\qquad 
\null [K, J_{0I}^{(n)}] &= \frac{n}{2} J_{0I}^{(n+1)} \, ,  \\
\null [H, J_{IJ}^{(n)}] &= n J_{IJ}^{(n-1)} \, , &\qquad 
\null [D, J_{IJ}^{(n)}] &=  n J_{IJ}^{(n)} \, , &\qquad 
\null [K, J_{IJ}^{(n)}] &= \frac{n}{2} J_{IJ}^{(n+1)} \, .
 \end{align}
 \end{subequations}
Due to the commutation relations and their differential operator realisation, we can understand $J_{IJ}^{(n)}$ as the generators of time dependent rotations. In contrast, $J^{(n)}_{0I}$ have commutation relations that resemble a momentum generator, but their differential realisation does not have the standard form.\footnote{To be in the standard form, their expression should be $J^{(n)}_{0I} = t^n S^0 \frac{\partial}{\partial S^I} - t^n S^I \frac{\partial}{\partial S^0}$.} Thus, we will call them time dependent pseudo-translations.
 
In addition to the symmetries we have found here, we could have included in our ansatz \eqref{ansatz_GED+scalars} a shift symmetry of the fields, e.g. $\delta \varphi = \text{eq.~\eqref{dphi_GED+scalars}} + f_\varphi(t,\vec{x})$, and similarly for the other fields. We find that these functions have to fulfil
\begin{subequations}\label{fix_func_2}
\begin{align}
\partial_i \partial^i f_{S^I}&=0 \, , &
\partial_t^2 f_\varphi +\partial_i \partial^i f_{\tilde{\varphi}}+\partial_t \partial_i f_{a_i} &=0 \, , \\
\partial_i \partial^i f_{\varphi}&=0 \, , &
\partial_i \partial^i f_{a_j} - \partial_j \partial^i f_{a_i} - \partial_t \partial_j f_{\varphi} &=0 \, .
 \end{align}
\end{subequations}

\section{Symmetries of the GYM action}
\label{sec:GYM_symmetries}

After analysing the case of GED with and without matter, we now move to study the symmetries of their non-abelian analogues. The GYM action in flat $3+1$ dimensions \cite{Bagchi:2022twx} is 
\begin{equation}\label{GYM}
S_{\text{GYM}}=\int \dd^{3}x\left[-\frac{1}{4}f^{ij\,a}f_{ij}\null^a- E^{i\,a}{\cal D}_i\varphi^a+\frac{1}{2}\left({\cal D}_t\varphi^a\right)^2\right]\,.
\end{equation}
where $f_{ij}\null^a$ and $E_i^a$ are defined similarly as before, but with an additional colour index structure 
\begin{equation}
f_{ij}\null^a\equiv \partial_i a_j{}^a-\partial _j a_i{}^a + \mathfrak{f}_{bc}{}^a  a_i{}^b a_j{}^c\, , \qquad
E_i\null^a\equiv {\cal \partial}_i\tilde\varphi^a-\partial_t a_i{}^a + \mathfrak{f}_{bc}{}^a a_i^b \tilde\varphi^c  \, . 
\end{equation}
The covariant derivatives ${\cal D}_i$ and ${\cal D}_t$ are defined as
\begin{equation}
{\cal D}_i \varphi^a =\partial_i \varphi^a + \mathfrak{f}_{bc}{}^a a_i^b \varphi^c\, , \qquad {\cal D}_t \varphi^a =\partial_t \varphi^a + \mathfrak{f}_{bc}{}^a \tilde\varphi^b \varphi^c\, , \label{def:covD}
\end{equation}
where $\mathfrak{f}_{bc}{}^{a}$ are the structure constants of the gauge group. As we are mostly interested in the $\mathfrak{su}(N)$ algebra, any repeated pair of colour indices is contracted with the Killing form.

We want to determine the spacetime symmetries of the GYM action. As these do not mix with the gauge transformations, the most generic ansatz is similar to \eqref{ansatz}, that is
\begin{eqnarray}
    \delta t = \xi^t \, , \qquad
    \delta x^i = \xi^i \, ,
\end{eqnarray}
and\footnote{The derivatives entering in \eqref{ansatz_GYM} are standard partial derivatives, as they appear from the Taylor expansion of the fields. }
\begin{subequations}\label{ansatz_GYM}
\begin{align}
\delta\varphi{}^a & =  \xi^t\partial_t\varphi{}^a+\xi^k\partial_k\varphi{}^a + \alpha_1 \varphi {}^a+ \alpha_2 \tilde{\varphi}{}^a + \beta^k a_k{}^a \,,\\
\delta\tilde\varphi{}^a & =  \xi^t\partial_t\tilde\varphi{}^a+\xi^k\partial_k\tilde\varphi{}^a + \alpha_3 \varphi{}^a + \alpha_4 \tilde{\varphi}{}^a + \gamma^k a_k{}^a \,,\\
\delta a_i{}^a & = \xi^t\partial_t a_i{}^a+\xi^k\partial_k a_i{}^a + \zeta_i \varphi{}^a + \varepsilon_i \tilde{\varphi}{}^a + \pi_i{}^k a_k{}^a \,,
 \end{align}
\end{subequations}
where $\xi^t, \xi^i, \alpha_i, \beta_i, \gamma_i, \zeta_i, \varepsilon_i, \pi_{ij}$ are all functions of the spacetime coordinates. We demand that the GYM Lagrangian transforms under (\ref{ansatz}) as a total derivative, i.e. $\delta \mathcal{L} = \partial_{\mu} F^{\mu}$, where $F^{\mu}$ is a generic function of fields, coordinates and parameters. This fixes the vector field components $\xi^t$ and $\xi^i$ as
\begin{subequations}\label{Xi_vector_GYM}
\begin{align}
\xi^t &= \frac{1}{2} \beta t^2 + \alpha t + \zeta \, , \\
\xi^i &= b_i(t) + \alpha x^i + \beta t x^i +  \lambda^i{}_k x^k \, ,   
 \end{align}
\end{subequations}
where $\alpha, \beta, \gamma, \zeta, \lambda_{ij}$ are constant parameters, with $\lambda_{ij} = - \lambda_{ji}$, and $b_i(t)$ are generic functions of $t$. The functions $\alpha_i, \beta_i, \gamma_i, \zeta_i, \varepsilon_i, \pi_{ij}$ are fixed in terms of derivatives of $\xi^t$ and $\xi^i$, which gives us
\begin{subequations}\label{func_fix_GYM}
\begin{align}
\alpha_1 &= \alpha_4= \alpha + \beta t \, ,  & \gamma_i &= -\zeta_i = \partial_t \xi_i= \beta x_i +\partial_t b_i
  \, ,  \\
 \pi_{ij} &= \partial_j \xi_i= \lambda_{ij} + \delta_{ij}(\alpha + \beta t) \, , & \alpha_2 &= \alpha_3 = \beta_i = \varepsilon_i =0  \, . 
 \end{align}
\end{subequations}
The symmetries of the GYM action generate an infinite dimensional algebra, spanned by the generators associated with the transformation parameters $\alpha, \beta, \zeta, \lambda_{ij}, b_i (t)$. They are:
\begin{subequations}\label{GYM_generators}
\begin{align}
\alpha&:  &D=t \partial_t +x^i\partial_i\, , \\
\beta&: &K=\frac{t^2}{2}\partial_t+tx^i\partial_i\, ,\\
\zeta&: &H=\partial_t\, ,\\
\lambda_{ij}&: &J_{ij}=x^i\partial_j-x^j\partial_i\, ,\\
b_i (t)&: &M^{(n)}_i =t^{n+1}\partial_i\, . 
 \end{align}
\end{subequations}
The non-vanishing commutation relations are exactly the same as \eqref{GED_symm_algebra} with $d=3$ and for the appropriate generator $D=D_t + D_x$, so we do not need to repeat them here. In contrast to GED, $D_t$ and $D_x$ are not symmetries of the GYM action by themselves, but only as their linear combination $D$. Therefore, the symmetries of the GYM action in $d=3$ are given by the \emph{Conformal Milne Algebra}.

\section{Symmetries of the GYM action with 5 adjoint scalars}
\label{sec:GYM_5scalars_symmetries}

Another non-relativistic limit of the $\mathcal{N}=4$ Yang-Mills action proposed in \cite{Fontanella:2024kyl, Fontanella:2024rvn} is GYM theory coupled to five extra scalar fields with quartic interaction that transform in the adjoint representation of the gauge group. In this section, we aim to identify the symmetries of this theory.

The GYM with scalars action in $d=3$ is the following
\begin{multline}\label{GYM_5scalars_action}
    S = S_{\text{GYM}} + S_{\text{scalars}} = \int \dd t \dd^{3}x \left[-\frac{1}{4}f^{ij\, a}f_{ij}{}^a - E^{i\, a} {\cal D}_i\varphi^a+\frac{1}{2}\left({\cal D}_t\varphi^a\right)^2 \right. \\
    \left.- \frac{1}{2} ({\cal D}_i S^{I\, a} )^2 - \mathfrak{f}_{ab}{}^c\varphi^a S^{I\, b} {\cal D}_t S^{I\, c} - \frac{1}{4} \left( \mathfrak{f}_{bc}{}^a S^{I\, b} S^{J\, c} \right)^2 \right]\, .
\end{multline}
where $S^{I\, a}$, with $I=1,..., 5$, are the five scalars. The covariant derivative of $S^{I\, a}$ is defined in the same way as for $\varphi^a$ in  \eqref{def:covD}.

Similarly to what we did for the GED theory, we can now ask whether the scalars $S^I$ break any of the GYM spacetime symmetries. To answer this question, we first write down the most general linear ansatz\footnote{Here, again there is no mixing between spacetime and gauge transformations.} 
\begin{subequations}\label{ansatz_GYM+scalars}
\begin{align}
\delta\varphi{}^a & =  \xi^t\partial_t\varphi{}^a+\xi^k\partial_k\varphi{}^a + \alpha_1 \varphi{}^a + \alpha_2 \tilde{\varphi}{}^a + \beta^k a_k{}^a + \tau_I S^{I\, a} \,,\\
\delta\tilde\varphi^a & =  \xi^t\partial_t\tilde\varphi{}^a+\xi^k\partial_k\tilde\varphi{}^a + \alpha_3 \varphi{}^a + \alpha_4 \tilde{\varphi}{}^a + \gamma^k a_k{}^a + \sigma_I S^{I\, a}\,,\\
\delta a_i{}^a & = \xi^t\partial_t a_i{}^a+\xi^k\partial_k a_i{}^a + \zeta_i \varphi{}^a + \varepsilon_i \tilde{\varphi}{}^a + \pi_i{}^k a_k{}^a + \chi_{iI} S^{I\, a}\,, \\
\delta S^{I\, a} & = \xi^t\partial_t S^{I\, a}+\xi^k\partial_k S^{I\, a} + \mu^I \varphi{}^a + \nu^I \tilde{\varphi}{}^a + \rho^{I k} a_k{}^a + \omega^I\null_{J} S^{J\, a} \, . 
 \end{align}
\end{subequations}
Demanding that the Lagrangian (\ref{GYM_5scalars_action}) transforms as a total derivative fixes the parameters $\xi^t, \xi^i, \alpha_i, \beta_i, \gamma_i, \zeta_i, \varepsilon_i, \pi_{ij}$ again as in (\ref{Xi_vector_GYM}) and (\ref{func_fix_GYM}). Therefore, the GYM symmetries are preserved. This procedure also fixes the remaining parameters $\mu^I, \nu^I,\rho^{I i},\omega^I\null_{J}, \tau_I, \sigma_I, \chi_{iI}$ to: 
\begin{subequations}\label{fix_func_2}
\begin{gather}
\mu_I = - \sigma_I = f_I(t) \, , \qquad
\omega_{IJ} = \tilde{\omega}_{IJ} +  \left(\alpha + \beta t  \right) \delta_{IJ} \, , \\
\nu_I = \tau_I = \chi_{iI} = \rho^{I i} = 0 \, .
 \end{gather}
\end{subequations}
where $\tilde{\omega}_{IJ}$ are antisymmetric, $\tilde{\omega}_{IJ} = - \tilde{\omega}_{JI} $, and $f_I(t)$ are generic functions of $t$. In contrast with the case of the GED with 5 scalars, now the rotations in the internal space $\tilde{\omega}_{IJ}$ cannot depend on time.

The terms associated with $\mu_I$ and the off-diagonal part of $\omega_{IJ}$ are purely internal symmetries, as it happens in the GED case. They give rise to the following generators,
\begin{subequations}
\begin{align}
f_I(t)&:  &J^{(n)}_{0I} = t^n \varphi \frac{\partial}{\partial S^I} - t^n S^I \frac{\partial}{\partial S^0} \, , \\
\tilde{\omega}_{IJ}&: &J_{IJ} = S^I \frac{\partial}{\partial S^J} - S^J \frac{\partial}{\partial S^I} \, . 
 \end{align}
\end{subequations}
where we again denoted $S^0 \equiv \tilde{\varphi}$. The non-vanishing commutation relations of these additional generators are exactly the same as \eqref{comm_rotations_GED} and \eqref{comm_HDK_rotations_GED} for the appropriate subset of generators, i.e. by identifying $J_{IJ}$ with $J_{IJ}^{(0)}$, so we do not need to repeat them here.\footnote{The full spacetime plus internal symmetries form an algebra of the type $\mathfrak{g} = \mathfrak{h}\oplus \mathfrak{f}$, where $[\mathfrak{h}, \mathfrak{h}] \subset\mathfrak{h}$, $[\mathfrak{h}, \mathfrak{f}] \subset\mathfrak{f}$, $[\mathfrak{f}, \mathfrak{f}] \subset\mathfrak{f}$, meaning that $\mathfrak{f}$ is an ideal of the algebra. Here $\mathfrak{h}$ is the conformal Milne algebra, whereas $\mathfrak{f}$ is the algebra of time-dependent pseudo-translations and time-independent rotations in $\mathbb{R}^5$, where the 5 scalars play the role of $\mathbb{R}^5$ coordinates.}

In addition to the symmetries we have found here, we could have included in our ansatz \eqref{ansatz_GYM+scalars} a shift symmetry of the fields. We find that the presence of the Yukawa-type coupling and the $S^4$ vertex forbids the presence of shift symmetries.

\section{An application to the non-relativistic AdS$_5$/CFT$_4$ correspondence}
\label{sec:application_NR_holography}

The gauge theory dual of the GGK non-relativistic string theory \cite{Gomis:2005pg} was studied in \cite{Fontanella:2024kyl, Fontanella:2024rvn}. As discussed in these papers, there are two possibilities for the boundary theory: one leading to GED with 5 free static scalars (i.e. a free theory), and the second one leading to GYM with 5 interacting adjoint scalars (i.e. an interacting theory). After the careful analysis of symmetries presented in this article, we conclude that the GED theory with 5 scalars \emph{cannot} be the dual of GGK non-relativistic string theory, because: I) it has a $D_x$ symmetry that is not realised in the bulk theory, II) it has a time-dependent rotation symmetry of the 5 scalars, whereas the bulk theory only admits these rotations to be time-independent, and III) it admits shift symmetries of the fields, which are not realised in the GGK theory.\footnote{The time dependent momentum symmetries of the GGK theory that emerge from flattening the 5-sphere are holographically realised in terms of $J_{0I}^{(n)}$. Although they appear with different differential realisation, they generate the same algebra.} 

On the other hand, the symmetries of the GYM with 5 scalars are the conformal Milne algebra plus the time-dependent pseudo-translations and time-independent rotations acting on the 5 scalars. These symmetries have been found in the bulk theory as Killing vectors evaluated at the boundary of the target space.\footnote{The Killing vector symmetries of the bulk theory include two copies of the Milne ``accelerations’’ $M_i^{(n)}$. This is because the bulk geometry has two longitudinal directions, $t$ and $z$. This means there is a tower of translations that depend on $t+z$ and another tower of translations that depend on $t-z$. These two copies of accelerations coincide at the boundary $z=0$, corresponding to $M_i^{(n)}$.} Therefore, this theory is the most likely holographic candidate out of the two.

\section{Conclusions}

We revisited the symmetries of the Galilean Electrodynamics (GED) action in a spacetime of generic dimension $d+1$. We found that they are always infinite dimensional and realised off-shell, in contrast to previous statements in \cite{Festuccia:2016caf,Bagchi:2022twx}. In the particular case of $d=3$, the symmetry is the conformal Milne algebra extended by the generator of spatial dilatations $D_x$. We found that adding 5 free static scalars to the action does not break any of these spacetime symmetries, and gives rise to time-dependent rotations and pseudo-translations of these scalars.

In the second part, we revisited the symmetries of the Galilean Yang-Mills (GYM) action, which we found to be the conformal Milne algebra, confirming the result in \cite{Bagchi:2022twx}. Then, we found that adding 5 scalars in the adjoint representation of the gauge group with the interaction given as in \cite{Fontanella:2024rvn} does not break the conformal Milne symmetry. In addition, we found that this action admits time-dependent pseudo-translations but time-independent rotations of the scalars. 

Finally, we discussed how the results of this article fit into the framework of non-relativistic AdS$_5$/CFT$_4$ correspondence \cite{Fontanella:2024kyl, Fontanella:2024rvn}. In particular, we commented that the symmetries of GYM with 5 adjoint scalars holographically match the symmetries of the GGK non-relativistic string theory action, therefore making this theory the potential holographic dual gauge theory. It would be interesting to use these symmetries to constrain the correlation functions of gauge invariant operators of the GYM with 5 scalars theory by imposing the Ward-Takahashi identities. Then, it would be useful to extract the corresponding conformal scaling dimensions and match them holographically with the string spectrum computed in \cite{deLeeuw:2024uaq}. This would be a highly non-trivial check, because the higher corrections in large string tension of the light-cone spectrum are all zero after complicated field redefinitions, as shown in \cite{deLeeuw:2024uaq}. Then, a similar situation is expected to happen in the dual GYM with 5 scalars, where the interaction terms should give a vanishing net contribution to the anomalous dimensions of dual operators.  We plan (and hope) to report on this in the future \cite{scaling_dim}.

\section*{Acknowledgments}
%%%%%%%%%%%%%%%%%%%%%%%%%%%%%%%%%%%%%%

We thank Neil Lambert for useful discussions. We thank the authors of \cite{Festuccia:2016caf} and \cite{Bagchi:2022twx} for private discussions about their results. We thank the referee Joseph Smith for pointing out that the symmetry generator $D_x$ of the GED theory disappears when a non-abelian structure is turned on.   
AF is supported by the SFI and the Royal Society under the grant number RFF$\backslash$EREF$\backslash$210373. JMNG is supported by the Deutsche Forschungsgemeinschaft (DFG, German Research Foundation) under Germany's Excellence Strategy -- EXC 2121 ``Quantum Universe'' -- 390833306 and by the Deutsche Forschungsgemeinschaft (DFG, German Research Foundation) – SFB-Geschäftszeichen 1624 – Projektnummer 506632645.
AF thanks Lia for her permanent support.

%%%%%%%%%%%%%%%% BIBLIOGRAPHY %%%%%%%%%%%%%%%%

\bibliographystyle{nb}

\bibliography{Biblio.bib}

\end{document}